\documentclass{emulateapj}

\newcommand{\kms}{km~s$^{-1}$}

\newcommand{\Ha}{$\rm H\alpha$}

\newcommand{\hi}{{H{\sc i}}}
\newcommand{\hii}{{H{\sc ii}}}
\newcommand{\nii}{\ion{N}{2}}
\newcommand{\oii}{[\ion{O}{2}]}

\newcommand{\bband}{{\em B}-band}
\newcommand{\iband}{{\em I}-band}
\newcommand{\rd}{$r_{\rm d}$}
\newcommand{\rmax}{$R_{\rm max}$}
\newcommand{\ropt}{$R_{\rm opt}$}
\newcommand{\rtw}{$R_{23.5}$}
\newcommand{\rpe}{$r_{\rm PE}$}
\newcommand{\vpe}{$V_{\rm PE}$}
\newcommand{\wpe}{$W_{\rm PE}$}
\newcommand{\whi}{$W_{21}$}
\newcommand{\whis}{$W_{80}$}
\newcommand{\wcou}{$W_{2.2}$}
\newcommand{\wpehis}{$W_{\rm PE}-W_{80}$}
\newcommand{\wpecou}{$W_{\rm PE}-W_{2.2}$}
\newcommand{\whipe}{$W_{\rm 21}-W_{PE}$}
\newcommand{\whicou}{$W_{\rm 21}-W_{2.2}$}
\newcommand{\sbo}{$\mu_0$}

\newcommand{\about}{$\sim$}

\slugcomment{Accepted for publication in the Astronomical Journal}

\shorttitle{Rotational Widths for TF Relation. II.}
\shortauthors{Catinella, Haynes, \& Giovanelli}

\begin{document}

\title{Rotational Widths for Use in the Tully-Fisher Relation. II. The Impact of
  Surface Brightness}

\author{Barbara Catinella\altaffilmark{1}, Martha P. Haynes\altaffilmark{2}, 
\& Riccardo Giovanelli\altaffilmark{2}}

\altaffiltext{1}{National Astronomy and Ionosphere Center, Arecibo Observatory, 
HC3 Box 53995, Arecibo, PR 00612, USA; bcatinel@naic.edu.
The National Astronomy and Ionosphere Center is operated by Cornell University 
under a cooperative agreement with the National Science Foundation.}
\altaffiltext{2}{Center for Radiophysics and Space Research and
  National Astronomy and Ionosphere Center, Cornell University,
  Ithaca, NY 14853, USA; haynes@astro.cornell.edu, riccardo@astro.cornell.edu.}

\begin{abstract}
Using a large sample of spiral galaxies for which 21 cm single-dish
and/or long-slit optical spectra are available, we make a detailed
comparison between various estimates of rotational widths. 
Different optical width estimators are considered and their
limitations discussed, with emphasis on biases associated with
rotation curve properties (shape and extent) and disk central
surface brightness.
The best match with \hi\ rotational velocities is obtained with Polyex
widths, which are measured at the optical radius (encompassing a fixed
fraction of the total light of the galaxy) from a model fit to the
rotation curve. In contrast with Polyex widths, optical rotational
velocities measured at 2.15 disk scale lengths \rd\ deviate from \hi\
widths by an amount that correlates with the central surface
brightness of the disk. 
This bias occurs because the rotation curves of galaxies are in
general still rising at 2.15 \rd, and the fraction of total mass
contained within this radius decreases with increasing disk surface
brightness.
Statistical corrections, parameterized by the radial
extent of the observed rotation curve, are provided to reduce Polyex and \hi\
width measurements into a homogeneous system. This yields a single
robust estimate of rotational velocity to be used for applications of
disk scaling relations.
\end{abstract}


\section{Introduction}

The importance of rotational velocities of spiral galaxies for
extragalactic astronomy and observational cosmology stems in large
part from their use in the Tully-Fisher \citep[TF;][]{tf77} relation. 
TF surveys initially relied on rotational velocities measured
from global (i.e., spatially integrated) \hi\ line profiles, typically
obtained using the Arecibo, Nan\c{c}ay, Green Bank, Parkes, and
Effelsberg radiotelescopes. Owing to sky coverage or sensitivity
limitations of single-dish antennas, optical long-slit spectroscopy
has been widely used to complement 21 cm \hi\ observations for TF
applications. The accuracy of the velocity width measurement is
particularly important for a TF survey, since the associated
uncertainties play a dominant role in the overall error budget of the
technique \citep[][hereafter G97b]{gio97b}.

The observed velocity width of a spiral galaxy spectrum is primarily
determined by the shape of its rotation curve (RC), the distribution
of the adopted kinematic tracer, the inclination of the plane of the
disk, and the measurement method. 
Integrated, single-dish \hi\ profiles provide very limited information 
regarding the distribution of the \hi\ gas, the shape of the RC, the 
presence of disk asymmetries, departures from planarity or non--circular
motions. However, the \hi\ gas is largely distributed in the outer 
parts of disks, sampling preferentially regions where the RC is flat 
or slowly rising; in addition, these measurements are completely 
independent of uncertainties in the determination of the disks' projected
major axis. Long-slit, major axis \Ha\ spectra yield an estimate of
the shape of the galaxy's RC, and they do so with substantial angular
resolution, allowing one to gauge departures from circular motion and 
disk asymmetries. Furthermore, they can probe the internal
kinematics of galaxies in high density environments, where the \hi\
signal could be {\em confused} by the presence of more than one target
within the telescope beam, or where the \hi\ emission could be truncated
because of interaction with the intracluster medium 
\citep[\hi\ {\em deficiency}; see, e.g.,][]{hg84,sol01}. Long-slit
optical RCs are also easier and faster to obtain at higher redshifts (i.e.,
$cz$ larger than a few thousand \kms) compared to \hi\ spectra.
However, they only sample the actively star forming  regions of the
disk, which typically extend only about half as far out as the
detectable \hi\ gas, and are sensitive to errors in the estimate 
of the major axis position angle, high internal extinction, and other
resolution-related effects (these issues are particularly important
for intermediate redshift galaxies; see, e.g., \citealt{kap06}). Thus
widths obtained with the two methods are differently affected by
systematic effects.

The cross-calibration of \Ha\ and \hi\ velocity width scales and the
understanding of the systematics involved in each technique are
essential for the combined use of heterogeneous data sets in TF
work. In fact, systematic differences between measures of rotational
velocities derived separately from optical and radio spectra could
mimic spurious bulk motions in peculiar velocity surveys, if the 
samples are segregated in the sky or in redshift space.
Such width conversion is also important for kinematic studies of
intermediate redshift ($z\gtrsim 0.4$) galaxies that must rely on
other tracers, such as \oii $\lambda$3727. Widths measured with these
diagnostics must be calibrated with those obtained with \hi\ and \Ha\
spectroscopy of local spirals \citep{kg00}.

As discussed in the first paper of this series \citep[][hereafter Paper I]{chg05}, 
the choice of the best algorithm to measure rotational velocities from
long-slit spectra is not an obvious one. Therefore here we compare
optical widths obtained from velocity histograms and from model fits
to the RCs and, for the latter, address the issue of the choice of the
reference radius for width measurement. In particular, we discuss the
vulnerability of optical widths to systematic effects associated with
RC shape and extent and disk central surface brightness (SB). For the RC
width estimator that is least affected by these biases, we provide the
cross-calibration with single-dish \hi\ widths. For a discussion of
how rotational velocities derived from global \hi\ profiles may differ
from those derived from \hi\ RCs, the reader is referred to, e.g.,
\citet{ver01}.
Comparisons between \hi\ and \Ha\ velocity widths have been previously
carried out by other authors, including \citet{mfb92}, \citet{ray97},
and \citet[][hereafter C97]{cou97}, based on overlap samples
substantially smaller than the one used here. However, the conversion
relations suggested by previous authors do not take into account the
biases discussed in this work. In particular, we show that the
cross-calibration of optical and radio widths depends on the shape and
extent of the observed RCs and, for widths measured at 2.15 disk scale
lengths (e.g., C97), on the SB of the galaxies.

In Paper I we presented the latest installment of \Ha\ spectroscopy for the SFI++ sample
\citep{smh07} and described our technique to measure rotational velocities from
long-slit spectra. We also showed that RC extent and \hi\ content are
correlated, in the sense that for a given optical size, galaxies with
larger \hi\ content are characterized by more extended RCs. This
result, which holds for objects with {\em normal} \hi\ content (i.e.,
not only for \hi-deficient, cluster members), will be used here to
explain observed trends in the comparison between optical and radio widths.
The average, or template, RCs of disk galaxies binned in luminosity
intervals obtained by \citet[hereafter CGH06]{cgh06}, together with
the SB effects on observed disk scale lengths and optical sizes
discussed therein, are also relevant for the analysis presented here.

The plan of this paper is as follows. The sample and velocity width
measurements from optical RCs and \hi-line profiles used in this work
are described in \S \ref{s_widths}. The comparison between optical and
radio rotational velocities is presented in \S
\ref{s_datacomp}. Systematic and SB effects are discussed in \S
\ref{s_disc} and our conclusions summarized in \S \ref{s_concl}.

\section{Data Sample and Velocity Width Measurement}\label{s_widths}

Our data set, referred to as SFI++, is a homogeneous compilation of
\iband\ photometry and long-slit \Ha\ and/or \hi\ rotational
parameters for $\sim$5000 galaxies, mostly late spirals.
The SFI++ catalog was designed for peculiar velocity studies in the
local universe using the TF method and is presented in \citet{smh07}.
Most of the SFI++ galaxies (85\%) have velocity in the CMB
frame cz$\leq$10,000 \kms\ (97\% have cz$\leq$15,000 \kms). Corrected
velocity widths range between 53 and 834 \kms\ (and between 200 and 500
\kms\ for 80\% of the objects), with a mean of 331 \kms.

The sample used in this work consists of 2753 SFI++ galaxies with high
quality long-slit \Ha/[\nii] RCs and \iband\ photometry, for which
good quality Polyex fits (see \S \ref{s_wopt}) are also available. 
Disk central SB values, \sbo\ (measured from the \iband\
profiles and corrected for Galactic and internal extinction,
cosmological k-term, and converted to face-on perspective), for this
sample vary between 14.0 and 21.3 mag arcsec$^{-2}$ (only two
objects have $\mu_0<14.0$ mag arcsec$^{-2}$); the distribution of
\sbo\ peaks at 18.6  mag arcsec$^{-2}$.
This data set was used for the comparisons involving optical widths
only. For those involving radio widths, we used the subset of galaxies
with high quality \hi\ spectra (uncertain or confused \hi\ profiles
were discarded, as well as a few outliers whose optical and radio
widths are clearly inconsistent, yielding a final sample of 873 objects).

\subsection{Optical Widths from Long-slit Rotation Curves}\label{s_wopt}

The algorithms commonly used to measure rotational widths from long-slit
spectra are based on either velocity histograms or functional fits to the
RCs. For the former, the width is obtained as the difference between
two percentile points of the RC velocity histogram; for the latter,
the width is measured at a fixed radial distance in the disk (obtained
from photometry) from the RC model fit.
The histogram technique is simple, resembles that used to measure
rotational widths from \hi\ integrated line profiles (although not
intensity weighted), and does not require accompanying photometry.
However, all the spatial information contained in the RC is lost and
systematic biases arise as a consequence.
In this work we characterize the biases affecting histogram width
determinations and compare model widths measured at two different radial
distances. 
We adopt the following RC width definitions:

\noindent
(a) $W_{80}=V_{90}-V_{10}$, the difference between the 90th and 10th
percentile points of the velocity histogram \citep[e.g.,][]{vog95}.
We determined $V_{10}$ and  $V_{90}$ for our RCs by linear
interpolation between the nearest velocity data points. This
definition is preferable over the difference between the maximum and
the minimum velocity of the RC \citep[e.g.,][]{mfb92,mf96}, because it
is less sensitive to the noise level of the RC (the maximum and the minimum
velocities typically correspond to the outermost detectable \hii\
regions, where the \Ha\ emission is usually fainter).

\noindent
(b) \wpe. This is our algorithm of choice for RC width measurement and
is described in detail in Paper I. In summary, we fold each RC
and fit the following function ({\em Polyex} model) to it:
\begin{equation}
        V_{\rm PE}(r) = V_0(1-e^{-r/r_{\rm PE}})(1+\alpha r/r_{\rm PE})
\label{eq_polyex}
\end{equation}
\noindent
where $V_0$, \rpe, and $\alpha$ determine the amplitude, the
exponential scale of the inner region, and the slope of the outer part
of the RC, respectively. The Polyex width is measured from the fit at
the optical radius
\footnote{
\ropt\ is defined as the radius encompassing 83\% of the total
integrated light. It is equivalent to 3.2 disk scale lengths for
a Freeman disk, i.e. an exponential disk with central surface brightness
$\mu_0 =21.65$ mag arcsec$^{-2}$ in \bband\ (corresponding to $\mu_0
\simeq 20$ mag arcsec$^{-2}$ in \iband, adopting an average $B-I$
color index of 1.7 from \citealt{dej96}).
}
\ropt: \wpe =2 \vpe (\ropt).
Other fitting functions have been proposed, either empirical
\citep[e.g., C97;][]{kkbp98,vog04} or based on physical models of disk
galaxies \citep[e.g.,][]{pss96}. The Polyex model has no physical
motivation, but has enough flexibility to fit the vast majority
(\about 94\%) of the individual RCs in our data set, as well as the
average RCs in luminosity intervals presented in CGH06. For $\alpha <0$,
it also allows one to fit declining RCs.

\noindent
(c) \wcou. We also discuss model widths measured from the Polyex fit
at 2.15 (hence \wcou) disk scale lengths \rd, another common choice of
reference radius for RC width measurement (see \S \ref{s_rdropt}).

Observed velocity widths derived from RCs must be corrected for
cosmological broadening (to obtain the rest-frame velocities) and
deprojected to an edge-on view:
\begin{equation}
        W^{\rm corr} = \frac{W^{\rm obs}}{(1+z)\,\sin i}
\label{eq_wcorr}
\end{equation}
\noindent
where $z$ and $i$ are redshift and inclination to the line of sight of
the galaxy, respectively. The uncertainties introduced by the
inclination correction are discussed in \citet[][hereafter G97a]{gio97a} and \citet{dal97}.
Corrections for turbulent (i.e., non-circular) motions of the gas,
usually applied to widths derived from \hi-line spectra (see \S \ref{s_whi}),
are considered negligible for RC widths. The reason is that, at any
spatial location along the slit, the RC is obtained from the velocity
of the peak of the line emission, and the broadening of the line due
to turbulent motions only affects its width (more details on RC
extraction can be found in Paper I).
Two other effects that could distort the observed RCs are slit
smearing (because of the finite width of the spectrograph slit,
regions that are slightly off the major axis could contribute to the
\Ha\ flux, smearing the RC) and slit misalignment. These effects
preferentially act in the sense of underestimating the true velocity
widths. Numerical simulations (to be presented elsewhere) show that
the combined effect of slit smearing and misalignment has a typical
amplitude of the order of 1-2\%, significantly smaller than the
accuracy of width measurements. Therefore we do not correct the
observed widths for these effects.

Velocity widths used in this work are corrected for cosmological
broadening {\em only}; the sin~$i$ correction is not applied in order
to minimize the uncertainties in our analysis, but we checked our
results for possible dependences on inclination. The plots shown in
this paper do not change significantly when the inclination-corrected
widths are used instead.

\subsubsection{Which Reference Radius for RC Width Measurement?}\label{s_rdropt}

The choice of the radius at which the model velocity widths should be
measured is subject to some debate. As discussed in Paper I, observed
RCs usually do not show a feature that could be easily associated with
the maximum circular velocity. In particular, RCs are in general flat
or even rising beyond 2.15 \rd, i.e. the location of the peak velocity
for a pure exponential disk \citep{fre70}. 
Compared to rotational velocities measured at 2.15 \rd, widths
evaluated at \ropt\ are less affected by extinction and are more
resilient, in the sense that a given error on the spatial scale
translates into a smaller scatter in the measured widths, because
velocity gradients are smaller in the outer disk. Moreover,
\ropt\ is a cumulative parameter, and hence more
straightforward to measure than \rd\ and less subject to extinction
effects. The use of \ropt\ avoids an additional complication. 
When the reference radius is a photometric
quantity, such as a multiple of \rd\ or an isophotal radius, its
location on the disk will vary with the inclination of the galaxy to
the line of sight. As a result, the corresponding measured width will
in general depend on inclination in a more complex form than purely
sinusoidal. The process could be standardized by using the radius at a
fixed inclination, e.g. face-on. However, the measured scale length
\rd\ of a given galaxy varies by about 50\% purely as the disk
inclination varies from intermediate to high values (see
e.g. \citealt{gio94}, Fig. 12), and the conversion 
\rd~$\rightarrow r^\circ_{\rm d}$ is quite uncertain. 
Thus, the reference radius should be chosen in the very outer disk,
where one may safely assume nearly total transparency and the
conversion to a face-on value is straightforward. 

One advantage of width measurements at 2.15 \rd\ is that most observed
RCs are mapped as far out as this radial distance, whereas only \about 60\%
reach \ropt. Thus, the measurement of \wpe\ at \ropt\ requires more frequently an
extrapolation of the fit. However, such extrapolation is mild, because, even if the RCs still
rise near \ropt, they do so at a small rate. Moreover, most of the RCs in our
sample (84\%) extend to at least 0.8 \ropt\ (only 2\% have extent
smaller than 0.5 \ropt. These are not used for TF applications).

For all the above reasons and in our experience, the total uncertainty
associated with the width measurement at \ropt\ is smaller than that
at 2.15 \rd. Ultimately, optical widths need to be compared to \hi\ ones, which
sample the galaxy kinematics out to larger radial distances in the
disks. These should therefore provide more reliable estimates of the
maximum rotational velocity, at least for objects in low density
environments, where the \hi\ distribution is undisturbed. Such a
comparison is presented in \S~\ref{s_hicomp}.

\subsection{Radio Widths from \hi-line Profiles}\label{s_whi}

Observations with single dish radio telescopes, whose beam areas
significantly exceed the angular size of the \hi\ distribution 
of a target galaxy, yield a spectrum of the galaxy's spatially
integrated \hi\ profile. Such profiles reveal neither the shape
of the RC nor the details of the \hi\ distribution, an argument that
has sometimes been used to suggest limited usefulness for single dish 21cm
widths (e.g. C97, \S 11). Nonetheless, such widths yield TF relations
of low scatter. This occurs because of the small slopes of RCs in the
outer parts of disks, where much of the gas tends to be located
(typically half of the \hi\ lies outside of about three scale lengths
from the center). This produces sharp outer edges to the spectral profiles,
so that a width measurement can be obtained with high precision
(generally $<$10 \kms). This circumstance is more significant
for luminous objects. For galaxies with rotational velocities of $\sim
75$ \kms\ or less, the outer slope of the RC is large and the
turbulent motions become relatively more important. As discussed in G97b,
the TF scatter among less luminous objects is quite large, partly
because the shape of the RC and the presence of turbulence make the derivation of
rotational velocity more uncertain, as are estimates of disk
inclination, but principally because the TF {\it intrinsic} scatter is
larger among them (see \citealt{hof96} as well as G97b). Therefore,
these objects are ill-suited as targets for TF studies for the
determination of distances or peculiar velocities, and such studies
may be better off without them. 

Rotational widths of disk galaxies are commonly obtained from
double-horned \hi\ profiles by measuring the velocity difference between
the outer N\% levels of each peak. Our technique is described in
detail in \citet{hay99b} and \citet[][hereafter S05]{shg05}. In summary, a straight
line is fitted to the receding side of the \hi\ profile, between the 15\%
and 85\% levels of the peak flux; $V_{\rm r}$ is the velocity for
which the line fit has a flux equal to 50\% of that of the peak. The
analogous process on the approaching side of the profile yields
$V_{\rm a}$; thus the measured width is \whi $= V_{\rm r} - V_{\rm a}$.

Standard corrections applied to \hi-line widths include those for instrumental
and cosmological broadening, turbulent motions, and deprojection to
edge-on view:
\begin{equation}
        W^{\rm corr} = \left(\frac{W^{\rm obs}-\Delta_{\rm s}}{1+z}
			-\Delta_{\rm t} \right) \frac{1}{\sin i}
\label{eq_whicorr}
\end{equation}
\noindent
The cosmological and inclination corrections are the same as those applied
to the optical widths (see eq. \ref{eq_wcorr}). The term
$\Delta_{\rm s}$ accounts for the effects of the spectrometer
resolution, the amount of smoothing applied, and the signal-to-noise
quality of the spectrum; the exact functional dependence of
$\Delta_{\rm s}$ on the latter two quantities must be modeled with
numerical simulations (S05). This correction is of the order
of the width of the velocity channels, typically between 5 and 11
\kms\ for our data. The correction for turbulent motions, 
$\Delta_{\rm t}$, is somewhat uncertain, and different values have been
adopted in the literature. S05 derived it for our data by
introducing a random, isotropic $\sigma =$10 \kms\ turbulent motion to
their simulations, and obtained $\Delta_{\rm t}=$6.5 \kms, independent
of profile width and signal-to-noise ratio.

As for the optical data, no inclination correction is applied to
the \hi\ widths. \\

For convenience, the definitions of all the radial distances and
rotational velocities adopted in this work are summarized in Table
\ref{t_definitions}.\\

\section{Velocity Width Comparisons}\label{s_datacomp}

\subsection{Optical Widths}\label{s_optcomp}

\begin{figure}
\epsscale{0.55}
\plotone{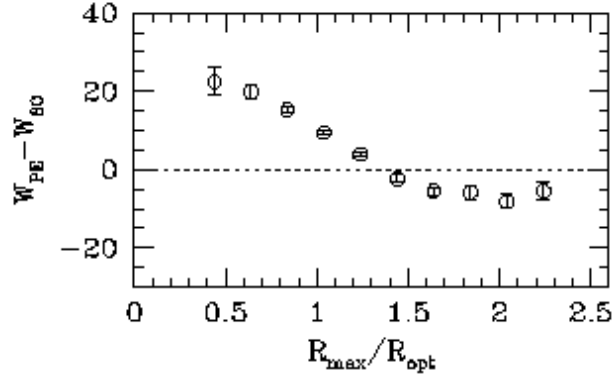}
\caption
{The difference between Polyex and histogram widths is plotted here as
  a function of RC extent. Data points are running averages; the bins
  are equally spaced, with separation 0.2 \rmax/\ropt; the error bars
  are Poisson errors on the mean. Velocity widths here and in all
  other figures of this paper are expressed in units of \kms. 
\label{opt_widths}}
\end{figure}

As mentioned in \S \ref{s_wopt}, width determinations based on
velocity histograms do not make use of the spatial information
contained in the RCs and this causes systematic deviations with
respect to measurements performed at a fixed radial scale.
On average, since the RCs are typically rising within the optical
disks (CGH06), histogram widths are expected to be systematically
smaller (larger) than \wpe\ for RCs less (more) extended than \ropt,
where the Polyex width is evaluated. This is shown in Figure
\ref{opt_widths}, where we plot the difference between Polyex and
histogram widths as a function of RC extent (expressed in units of
\ropt). The effect is smaller for RCs with larger extent, because
galaxies with intrinsically large \Ha\ disks are also more luminous, and
therefore their RCs tend to be flatter (CGH06). 

\begin{figure}
\epsscale{1.0}
\plotone{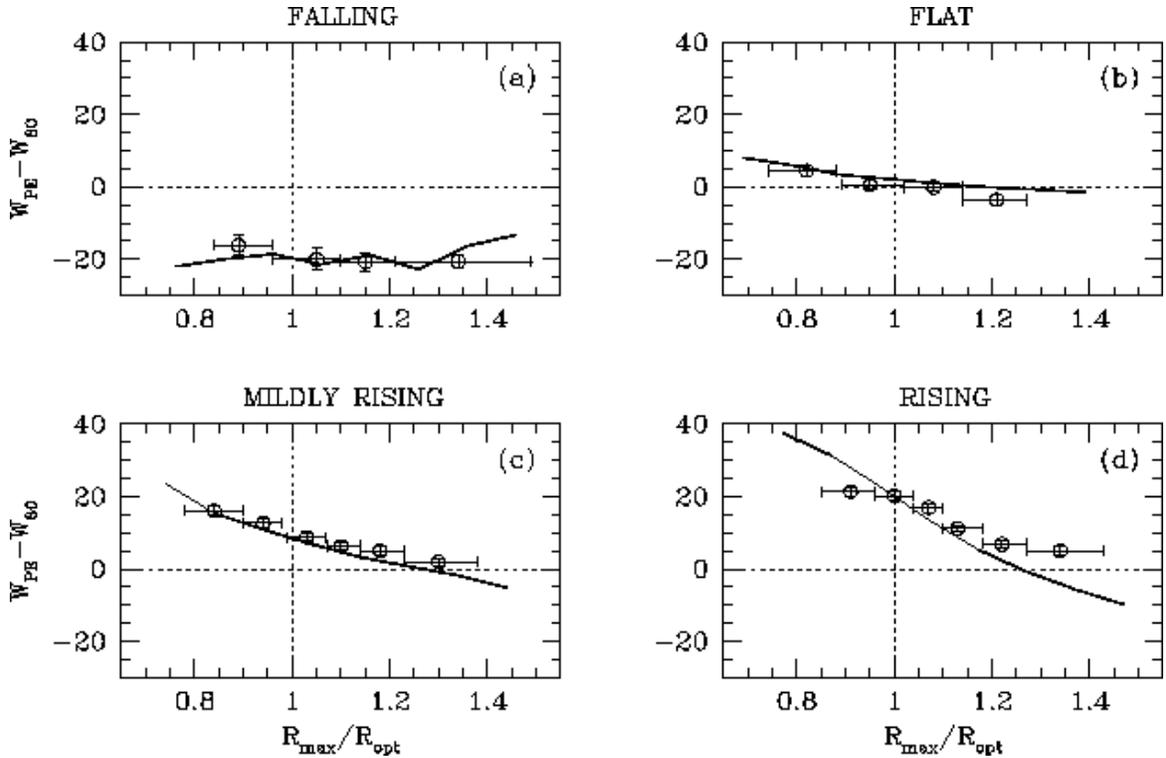}
\caption
{The width difference in Figure \ref{opt_widths} is plotted here for 4
  classes of increasing RC slope: 
  (a) $\theta<0$; (b) $0 \leq \theta<0.5$; (c) $0.5 \leq \theta<1.2$;
  (d) $\theta \geq 1.2$ (where $\theta$ is the slope of the Polyex
  fit between 0.5 and 1.0 \ropt; its units are \kms\ arcsec$^{-1}$).
  The data points are running averages, each including 25, 100, 100,
  and 150 measurements in panels (a) through (d), respectively;
  vertical and horizontal error bars indicate Poisson errors on
  the mean and bin sizes, respectively. The data lying outside the spatial range
  displayed are sparse and are not shown. The vertical, dotted line is
  at \rmax=\ropt, where \wpe\ is measured (for RCs to the left of this
  line, the measurement of \wpe\ requires an extrapolation of the fit).
  The solid lines show the results of numerical simulations, as
  discussed in the text.
\label{widths_sim}}
\end{figure}

In order to
characterize the dependence of histogram width measurements on the
velocity gradient of the RCs, we divided our sample into four bins
with increasing RC outer slope, as obtained from the Polyex fit
between 0.5 and 1.0 \ropt. Figure \ref{widths_sim} shows the
difference between \wpe\ and \whis\ as a function of RC extent for the
four bins, roughly corresponding to falling (a), flat (b), mildly
rising (c), and  rising (d) RCs. Several features can be noticed:
(1) the difference \wpehis\ depends on both extent and slope of the RC;
(2) \whis\ and \wpe\ yield the same results {\em only} for flat RCs,
traced out to a radial distance of the order of \ropt;
(3) over the \rmax/\ropt\ range considered here, \whis\ is smaller than
\wpe\ for rising RCs, the difference increasing for RCs with steeper
outer slope and smaller extent. In the case of falling RCs, \whis\ is
larger than \wpe;
(4) the intercept of the relation with the axis \rmax/\ropt=1 increases 
from approximately $-$20 to 20 \kms\ from falling to rising RCs.

We performed numerical simulations to model the trends observed in
Figure \ref{widths_sim}. We simulated mock catalogs of galaxies with
ideal, perfectly symmetric and noiseless RCs, assigning RC extent
(\rmax/\ropt) and Polyex fitting parameters with the same frequency
distributions observed in the SFI++ sample (paying attention to the
fact that the distributions of the RC parameters are {\em not}
independent from each other). We calculated RC slopes and widths, and
binned the simulated data as done for our observations. The results
are shown as solid lines in Figure \ref{widths_sim}. Even with the
simplistic RC description adopted, our simulations are able to
reproduce the systematic trends observed in the data. A more realistic
modeling, taking into account statistical distributions of \Ha\
emission extent, RC asymmetry, and measurement errors, could easily
explain the remaining (small) differences between the data and the
results of the simulations presented here. These results demonstrate
that the systematic biases affecting histogram width measurements are
completely explained, once the extents and shapes of the RCs are taken
into account. RC widths based on velocity histograms will not be
further discussed. 

The next issue we want to address is the choice of the reference
radius at which model widths should be evaluated, \ropt\ or 2.15 \rd. 
The uncertainties affecting the measurement of \ropt\ and \rd\ have
been discussed in \S \ref{s_rdropt}; in this section and in 
\S \ref{s_hicomp} we will compare \wpe\ and \wcou\ with each other and
with velocity widths derived from \hi\ spectra.

The mean width difference \wpecou\ for our sample is 17 \kms\ 
(with a dispersion $\sigma$=19 \kms). This offset is expected from
the fact that, on average, the RCs are still rising at 2.15 \rd\
(CGH06, Fig. 1) and \wpe\ is measured farther out in the
disk. However, values of \wpecou\ for individual galaxies show a large
variation, from $-$32 to over 100 \kms\ (see Fig.~\ref{opt_widths_sb} below).

As opposed to histogram widths, model widths do not depend on RC
extent, as long as the RC is sampled to a radius large enough to
warrant a reliable fit. In fact, \wpecou\ does not show systematic
variations when plotted as a function of \rmax/\ropt\ (not shown).
On the other hand, the difference between widths measured
at \ropt\ and 2.15 \rd\ should increase for larger \ropt/\rd\
ratios, and thus should depend on SB, because galaxies with larger
values of \ropt/\rd\ are typically of higher optical SB (CGH06).

In order to quantify these biases and gain more
insights on the role of the SB of a galaxy, we studied the variation of
the width difference as a function of the \ropt/\rd\ and \rtw/\rd\ ratios
\footnote{
\rtw\ is the isophotal
radius measured at an \iband\ surface brightness of 
23.5 mag~arcsec$^{-2}$. This approximately corresponds to $R_{25}$ in
\bband, assuming an average $B-I$ color index of 1.7 \citep[from][]{dej96}.
}
and of the \iband\ disk central SB, \sbo.
Figure \ref{opt_widths_sb}a shows the dependence of \wpecou\ on the
\ropt/\rd\ ratio for the whole sample. This correlation is entirely
expected from the fact that the RCs are typically still rising at 2.15 \rd,
where \wcou\ is measured. Therefore, for a given \ropt, \wpecou\
will increase for smaller values of \rd, i.e. larger \ropt/\rd\
ratios, as observed. The bins corresponding to larger width
differences have larger error bars, due to both smaller
number statistics and larger scatter among the individual data that
were combined. 

As already mentioned, galaxies of higher \sbo\ have typically larger
values of \ropt/\rd\ and thus, based on Figure \ref{opt_widths_sb}a,
they should also have larger \wpecou\ width differences. This is
confirmed in panel (b), where the sample has been divided into three
intervals of central SB (corresponding to those adopted in Fig. 9 of
CGH06, except that the two central bins in that figure have now been
combined). 

It is perhaps more insightful to study the dependence of the width
difference on the \rtw/\rd\ ratio, since an isophotal radius has a
more straightforward dependence on the disk central SB than \ropt.
In fact, the observed brightness profile of an exponential disk,
expressed in magnitude terms, can be written as:
\begin{displaymath}
        \mu (r) = \mu_0 + 1.086 ~r/r_{\rm d}
\end{displaymath}

\noindent
which implies a linear relationship between \sbo\ and the \rtw/\rd\ ratio:
\begin{displaymath}
        \mu_0 = 23.5 - 1.086 ~R_{23.5}/r_{\rm d}
\end{displaymath}
Since \ropt\ and \rtw\ are clearly correlated, this also implies a
correlation between \sbo\ and \ropt/\rd, but the scatter is
of course larger. To this extent, it is worth remembering that \ropt\ is defined
as the radius encompassing a fixed fraction of the total light of the
galaxy, i.e. it is not a pure disk quantity. Therefore, its value
depends on the degree of
concentration of the light as well as on the possible presence of a
spheroidal component. For a given total luminosity, smaller
exponential disks clearly have larger \ropt/\rd\ ratios. 
For a Freeman disk, \ropt/\rd $=3.2$ (see footnote 1); galaxies
of higher SB and similar values of \rd\ have larger \ropt/\rd\ ratios.
However, notice that a galaxy harboring a bulge has a {\em smaller}
\ropt/\rd\ than a galaxy with the same exponential disk but no bulge. 
Figures \ref{opt_widths_sb}c and \ref{opt_widths_sb}d 
illustrate the behavior of the width difference as a function of 
\rtw/\rd\ for the whole sample and separately for the three SB
intervals, respectively. Panel (d) shows that galaxies of different SB
are well segregated in the \rtw/\rd\ parameter space and that {\it larger
width differences are clearly associated with higher disk central SB}.
This will be further discussed in \S \ref{s_disc}.

In order to decide which model width, \wpe\ or \wcou, is best suited
for use in the TF relation, we will compare these with width
measurements obtained from \hi\ spectra in \S \ref{s_hicomp}. The
optical width of choice will be that providing the best match to
\whi\ based on the amount of offset and scatter of the correlation
and, most importantly, vulnerability to systematic effects.

\subsection{Optical vs. Radio Widths}\label{s_hicomp}

The results presented in the previous section showed that the
difference between optical model widths measured at \ropt\ and 2.15
\rd\ depends on the SB of the galaxy. Thus we first evaluate the impact of
SB-related biases on comparisons between radio and optical rotational
velocities by plotting the corresponding width differences as 
functions of the \rtw/\rd\ ratio. As Figure \ref{hiwidths_sbr235}
clearly demonstrates, \wpe\ provides a more reliable estimate of the
full width of disk galaxies, as measured from spatially integrated
\hi-line profiles, than does \wcou. In fact, not only is the difference
between \whi\ and \wpe\ always smaller than that between \whi\ and
\wcou, but it is also not affected by systematic trends, whereas
$W_{\rm 21}-W_{2.2}$ increases for larger \rtw/\rd\ ratios,
i.e. galaxies of higher SB. Plotting the width differences as
functions of the central SB yields analogous results.
Therefore we will only cross-calibrate \hi\ and
Polyex widths and discuss the implications of the SB effects
associated with RC widths measured at 2.15 \rd\ in \S \ref{s_disc}. 

As seen in Figure \ref{hiwidths_sbr235}, \hi\ widths are
systematically larger than \Ha\ ones; the average value of \whipe\
is $14 \pm 1$ \kms\ (with a dispersion $\sigma$=29 \kms).
Interestingly, the width difference tends to increase for galaxies
with extended RCs, as shown in Figure \ref{radio_widths}a. The
velocity increase is small for the whole sample and the scatter is
large (excluding the two bins with \rmax/\ropt $\geq$2,
which include less than 20 objects each, the velocity dispersion
increases monotonically from 20 to 53 \kms\ with decreasing RC extent),
nonetheless the effect is systematic. 
The same bias can be seen in panel (b), where the
ratio, rather than the difference, of the two widths is plotted.
The solid lines in panels (a) and (b) are
quadratic fits to the data; their analytical expressions can be found
in the first two rows of Table \ref{t_wconv}, respectively.

The fact that the RCs might still be rising beyond \ropt\ could explain the
offset between \whi\ and \wpe\ (as long as \hi\ emission is detected
beyond \ropt), but not the dependence of their ratio on RC
extent. The latter requires that RC and \hi\ extents are also
correlated. In Paper I we showed that, for a given optical size, star formation can be traced
farther out in disks of galaxies with larger \hi\ content. We also
pointed out that this result is not limited to the well known case of
truncated star formation in \hi-poor cluster objects, but also holds for
\hi{\em-normal}, field spirals. From the analysis of a sample of 108
\hi-rich galaxies with available \hi\ synthesis maps,
\citet{br97} showed that \hi\ mass and \hi\ diameter are tightly
correlated; moreover the \hi\ diameter correlates well with the
optical size of the galaxy. Therefore, a correlation between RC extent
and \hi\ size seems to be a reasonable assumption. 
Thus, biases might arise because for galaxies with larger \Ha\ extent
the \hi\ width traces farther out into the halo, while we continue to
measure \wpe\ at \ropt.
This also implies that the dependence of the width ratio (or
difference) on RC extent should be larger for rising RCs and
negligible for flat ones. This is clearly demonstrated in Figure
\ref{awph_wph_slp}, where the sample is divided into two classes with
different RC slopes at the optical radius. The dependence of
\whi/\wpe\ on RC extent is present for galaxies with rising RCs
(a), but effectively disappears for the rest of the sample (b). Linear
fits to the data points are shown as solid lines and listed in the
last two rows of Table \ref{t_wconv}. 
Notice that, for rising RCs with extent smaller than \ropt, \wpe\
tends to be larger than \whi, implying that the extrapolation of the RC to
\ropt\ (where \wpe\ is measured) often overestimates the width
measured from \hi\ profiles. However, it should also be noticed that the
first bin of panel (a) includes only 20 elements and that the scatter in
these plots is large (the value of the 1$\sigma$ dispersion around the
mean is 0.15 for the first bin and \about 0.10 for the other bins of
panel [a], and varies between 0.07 and 0.20 for panel [b]).

The common assumption that optical widths underestimate radio
ones, because RCs are often rising and the \hi\ gas extends farther out
in the disk compared to \Ha\ emission, is confirmed by our data. As
argued above, the observed dependence of \whi/\wpe\ on RC extent is
also in agreement with this picture, if one assumes that galaxies with
larger RC extent also have larger \hi\ size. 
However, these arguments do not explain why \hi\
widths are systematically larger than \Ha\ ones for galaxies with 
{\em flat} RCs (Fig. \ref{awph_wph_slp}b). 
Even adopting a much stricter definition of flat RC (i.e., 
$-0.1<\Theta <0.1$, a requirement met by 161 objects) yields an
average velocity offset \whipe\ of $17 \pm 2$ \kms.
At least part of this offset might indicate that the turbulent motion
correction applied to radio widths (6.5 \kms, see \S~\ref{s_whi}) is
underestimated. Not only is such a correction somewhat uncertain, with
various values being advocated by different authors, but also it
does not take into account tidal disturbances, bars, and non-planar
motions which can occur frequently and with varying amplitude.
Another important factor to keep in mind is that widths obtained from
global \hi\ profiles depend on both RC shape and distribution of the
\hi\ gas. 
Resolved \hi\ RCs of nearby galaxies show that, for most intermediate
and bright objects, a velocity peak is often reached before or
near \about 3 \rd, followed by a small decline and a flat outer part
(e.g., \citealt{cvg91}; \citealt{bro92}; C97; \citealt{ver01}). 
In these cases, the width of the global \hi\ profile measures
the peak velocity, whereas the Polyex width obtained from the \Ha\ RC
would yield the amplitude of the flat part (as typically \ropt/\rd
$>$3.2 for high SB galaxies).

The results presented in this section indicate that, in order to
cross-calibrate optical and radio rotational velocities, both extent
and shape of the RCs should be taken into account. The relations
presented in Table \ref{t_wconv} allow us to combine Polyex and \hi\
velocity widths into a homogeneous data set, which can be used for
applications of the TF distance method without introducing significant
systematic biases.

\section{Discussion}\label{s_disc}

\subsection{Systematic Effects}\label{s_syst}

The comparison between \hi\ and \Ha\ velocity widths has been carried
out by several authors. \citet{cou92}, \citet{mfb92}, \citet{vog95}, and
\citet{ray97} found tight correlations between radio and optical
histogram widths, with slopes approximately equal to unity, small zero
points ($\leq 5$ \kms), and dispersions ranging between 10 and 30 \kms.
The most comprehensive calibration to date is that of C97, who
compared optical widths measured with several algorithms for a sample
of 304 galaxies with \hi\ widths obtained by G97a (we note that the G97a
data set is part of the SFI++ sample, but the \hi\ widths have
recently been recomputed using different instrumental and turbulent
motion corrections, following S05). 
He concluded that the best match to 21 cm widths is obtained with
model optical widths measured at 2.15 \rd\ (using a fitting function
different from the Polyex model adopted here). This was based on the observation
that the other RC widths examined in his work (including histogram
types and model widths measured at 3.2 \rd) yield correlations with
radio widths having larger zero points and scatters (but similar slopes). 

However, direct comparisons between different width estimators are not
the best way to establish the most suitable algorithm for
optical width measurement, because they are not sensitive to the
presence of systematic biases in the data. In this paper, we argue
that model \Ha\ widths measured at the optical radius (defined as a
cumulative quantity and {\em not} as a fixed multiple of exponential
disk scale lengths) provide the best match to \hi\ widths because they
are {\em the least affected by systematic effects}. In fact, we have
demonstrated that histogram widths deviate with respect to measurements
performed at a fixed radial scale in a systematic and predictable way,
depending on the shape and extent of the RCs. Next, we have compared
model widths measured at \ropt\ and 2.15 \rd, \wpe\ and \wcou, and
shown that their difference is larger for galaxies with higher
disk central SB. The comparison with \hi\ rotational velocities indicates
that \wpe\ measurements are not affected by SB effects, contrary to
\wcou\ ones. As shown in Figure \ref{hiwidths_sbr235}, the difference
between \hi\ and Polyex widths plotted as a function of the \rtw/\rd\
ratio shows no systematic trends, whereas \whicou\ increases for
larger \rtw/\rd\ values (i.e. higher SB objects), varying from 
\about 20 to \about 60 \kms. 
However, even Polyex widths are not completely immune to systematic
effects, as the \whipe\ difference (or their ratio) still shows a
slight dependence on the radial extent of the observed RC, for which we
provide a statistical correction. Such dependence also exists for
histogram and \wcou\ widths (see \citealt{cat05}).

Systematic effects such as those mentioned above
must be taken into account in order to avoid biases in the analysis
of TF surveys. As mentioned in Paper I, a width difference
of 20 \kms\ translates into approximately a 0.2 mag offset on the TF
plane, or 500 \kms\ in peculiar velocity, at $cz\sim 5000$ \kms. 
Thus spurious bulk flows of significant size can arise when 
heterogeneous data sets (e.g., optical and radio) are combined that do
not share similar distribution properties, such as in sky projection
or redshift. The same considerations apply to a homogeneous sample of
galaxies spanning a range of central SB and redshift, if \wcou\ is
used as width estimator. In fact, higher redshift bins would
preferentially include higher SB objects and these require larger
width corrections, as seen in Figure \ref{hiwidths_sbr235}b.

\subsection{Surface Brightness}\label{s_sb}

In this section we discuss in more detail the impact of galaxy
SB and RC shape on optical width measurements. 

Velocity widths obtained from model fits to the RCs depend on both 
RC shape and radius at which the measurement is performed.
When model widths measured at 2.15 \rd\ and \ropt\ are compared, a
dependence on disk central SB is also observed.
This was shown in Figure \ref{opt_widths_sb} and is illustrated
more clearly in Figure \ref{widths_sb}, where \wcou/\wpe\ is
plotted as a function of the \iband\ disk central SB, \sbo.  
The width ratio is approximately constant for low SB (LSB) galaxies\footnote{
We use ``LSB'' to indicate the lower end of the
SB interval spanned by our data. However, it should be noted that our
sample includes very few truly low SB systems (see \S~\ref{s_widths}).
}  
and decreases for high SB (HSB) ones. The reason for this trend is
that HSB galaxies have larger \ropt/\rd\ ratios. Since the RCs are
typically rising at 2.15 \rd, this implies that \wcou\
reaches a smaller fraction of the velocity at \ropt\ in HSB galaxies.

Observed rotational widths yield information on the total (i.e., baryonic
and dark) mass contained within the radius where they are measured.
In particular, \wcou\ is measured at the radius where the RC of
the exponential, stellar disk would peak (i.e., where the contribution
of the baryonic component to the total velocity field is maximized),
whereas \wpe\ is measured at a larger radius,
where the contribution of the dark matter to the RC is expected to be
larger. The dependence of the width ratio on SB seen in Figure
\ref{widths_sb} indicates that the fraction of total mass contained within 2.15
\rd\ decreases for galaxies of higher disk central SB.
This yields support to the commonly accepted notion that in HSB
galaxies the dark matter contribution becomes important only at large
radii, whereas LSB galaxies are already dark matter dominated at small
radii (e.g., \citealt{cvg91}; \citealt{db02}). In
particular, the severity of the mass discrepancy correlates with
central SB, being larger for lower SB (e.g., \citealt{mcg98} and
references therein). 

Since RC shapes are known to depend strongly on luminosity (e.g., 
\citealt{rub85}; \citealt{pss96}; CGH06), it is interesting
to look at the dependence of the \wcou/\wpe\ ratio on both central SB 
and luminosity. 
Figure \ref{widths_msb} shows how the ratio between the two model widths
varies with \iband\ absolute magnitude, for galaxies of different central
SB. The three SB classes correspond to
those defined in Figure \ref{opt_widths_sb} (namely, the HSB/MSB and
MSB/LSB divisions occur at \sbo\ of 17.5 and 19.5 mag~arcsec$^{-2}$,
respectively). Since the shape of the RCs depends on luminosity and the
\ropt/\rd\ ratio depends on \sbo, this plot clarifies the interplay
between RC shape and choice of reference radius when one measures
velocity widths from RCs. Typically, galaxies on the left (right) hand
side of the plot have flat (rising) RCs, and the \ropt/\rd\ ratio
increases from top (LSB) to bottom (HSB). For a given luminosity,
the \wcou/\wpe\ width ratio is smaller for HSB galaxies, because the RC shape is
approximately fixed but the separation between the radii where the
widths are measured increases. Conversely, along a ``constant'' SB
line the \ropt/\rd\ ratio does not vary strongly, but the RC shape
changes substantially, and larger width discrepancies are associated with rising
RCs. \wcou\ and \wpe\ show the best agreement for high luminosity, LSB
galaxies and the largest discrepancy for lower luminosity, HSB
objects.

As mentioned above, the shapes of the RCs are known to depend strongly on
luminosity but also, as numerous authors have pointed out, on SB.
Disentangling the relative importance of these two quantities for the
shapes of the observed RCs of galaxies spanning a wide range of
properties is not a trivial task. Nonetheless, this would be helpful
to improve our understanding of the systematic trends affecting
optical width measurements discussed here.
A detailed study of how RC shapes vary with the SB of a 
galaxy for the SFI++ sample is beyond the scope of this paper and
will be explored in a future work.

\section{Summary and Conclusions}\label{s_concl}

In this work we have made a detailed comparison of rotational widths
obtained from long-slit optical and single-dish 21 cm spectra.
Because the combination of heterogeneous data sets is necessary for
the study of many astrophysical problems, this comparison is important
in underscoring potential sources of bias. 

Since the choice of the best RC width for TF applications is somewhat
debated, we have considered different RC width estimators and
discussed their limitations. In particular, we have studied their
vulnerability to biases related to the shape and extent of the RCs and
to the disk central SB of the galaxies. Our main conclusions are:

(a) Optical rotational widths obtained from RC velocity histograms are
affected by systematic biases that correlate with the shape and the
extent of the RCs. The systematics observed in the data are well
reproduced by our numerical simulations.

(b) A more reliable width estimate that makes use of the spatial
information contained in the RCs is obtained by fitting a function
to the RC and measuring the velocity at a fixed radial distance from
the center of the disk. We refer to these as {\em model} widths.
In this work, we used the Polyex model (eq. \ref{eq_polyex})
to fit the RCs. Velocity widths measured from the RC fit at
the optical radius, \ropt, are referred to as {\em Polyex widths}, \wpe.

(c) The comparison between model widths measured at \ropt\ and
2.15 \rd, \wpe\ and \wcou, shows a dependence on 
disk central SB. Larger width differences are observed for
galaxies of higher SB, which are characterized by larger \ropt/\rd\
ratios. For a fixed SB, the \wcou/\wpe\ ratio decreases for galaxies
of lower luminosity, which typically have rising RCs.
Thus this bias occurs because RCs are in general still rising at 2.15 \rd,
and the fraction of total mass included within that radius is smaller
for HSB galaxies.

(d) The best match with \hi\ rotational velocities is obtained with
RC widths measured at the optical radius. In fact, \whipe\ is always
smaller than \whicou\ and, most importantly, is not affected by SB
effects, whereas \whicou\ is larger for galaxies of higher disk central SB.

(e) When compared with 21 cm measurements, all the optical widths
examined in this work show a dependence on the radial extent to which
RCs are measured. For the model widths, larger offsets from \whi\ are
observed for galaxies with \Ha\ emission traced farther out in the
disk. If the radial extent of the \hi\ emission and that of the
detectable \hii\ regions are correlated (see \S~\ref{s_hicomp}), galaxies with
larger \Ha\ disks and rising RCs should also have larger \hi\
extents and widths. In fact, we have shown that the dependence of
\whi/\wpe\ on RC extent exists only for rising RCs and effectively
disappears for flat ones.

Lastly, we have provided statistical corrections, parameterized by the
radial extent of the observed optical RC, that can be applied to
Polyex measurements in order to approximate consistency with \hi\
rotational widths. Such cross-calibration has been used for the
derivation of a TF template relation for the SFI++ sample
\citep{mas06}.
Failure to correct observed rotational widths for the systematic
effects discussed here will produce biases in astrophysical inferences,
most notably in the measurement of the peculiar velocity field and the
study of the evolution of galactic mass--to--light ratios. These
biases can become particularly significant when data sets are combined
that do not share similar distribution properties, such as in sky
projection or redshift.

\acknowledgements

This work has been partially supported by NSF grants AST-0307396 and
AST-0307661 to MPH. BC thanks Chris Salter for helpful comments. 
We also wish to thank our anonymous referee for constructive
comments which helped us to improve this paper.

\clearpage

\begin{figure}
\epsscale{1.0}
\plotone{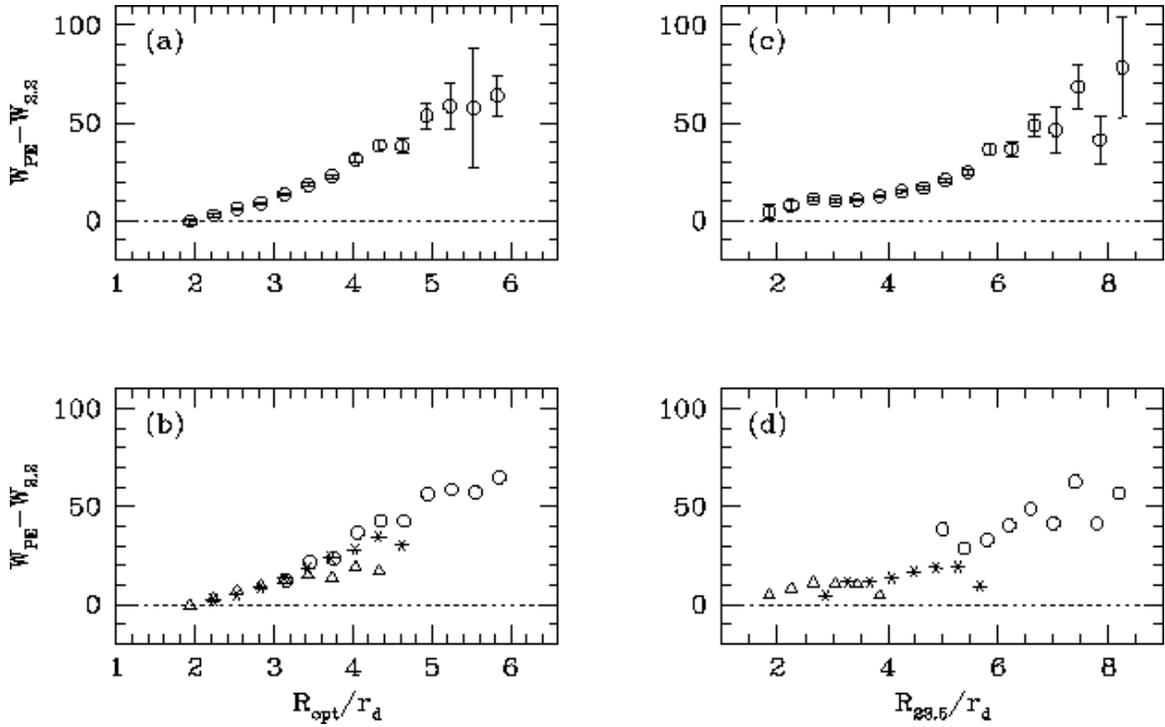}
\caption
{The difference between model widths measured at \ropt\ and 2.15 \rd\
  is plotted here as a function of \ropt/\rd\ (left panels) and
  \rtw/\rd\ (right) ratios. In the lower panels, the data are divided
  in 3 classes according to the galaxy's \iband\ disk central surface brightness:
  $\mu_0 \leq 17.5$ (circles), $17.5 < \mu_0 \leq 19.5$ (asterisks),
  and $\mu_0 > 19.5$ mag~arcsec$^{-2}$ (triangles). In all panels,
  data points are bin averages; the errors on the mean are not shown
  in (b) and (d).
\label{opt_widths_sb}}
\end{figure}

\begin{figure}
\epsscale{0.55}
\plotone{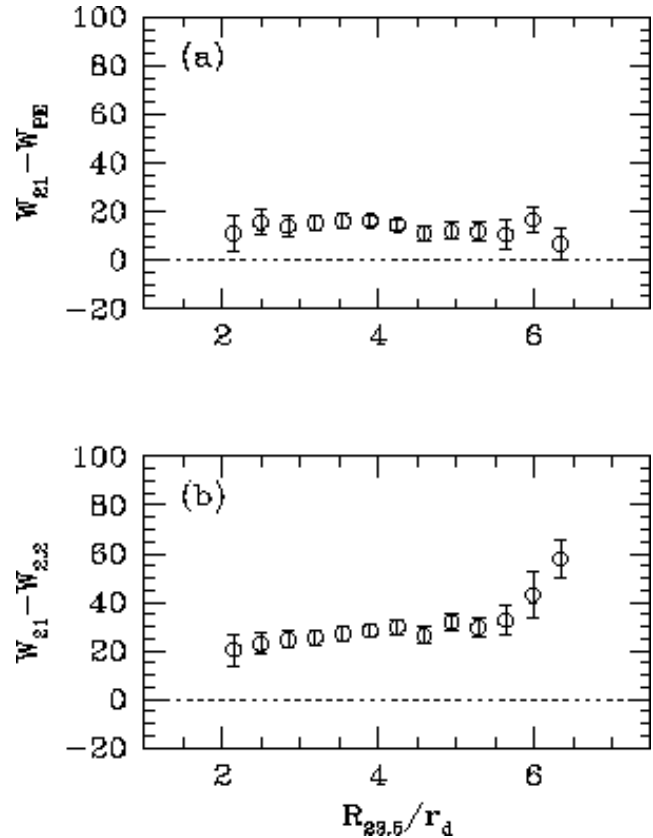}
\caption
{(a) Polyex widths are compared to \hi\ measurements as a function of
  the \rtw/\rd\ ratio. (b) Same as (a) for RC model widths measured at
  2.15 \rd. Data points and error bars represent running averages and
  errors on the mean, respectively.
\label{hiwidths_sbr235}}
\end{figure}

\begin{figure}
\epsscale{0.55}
\plotone{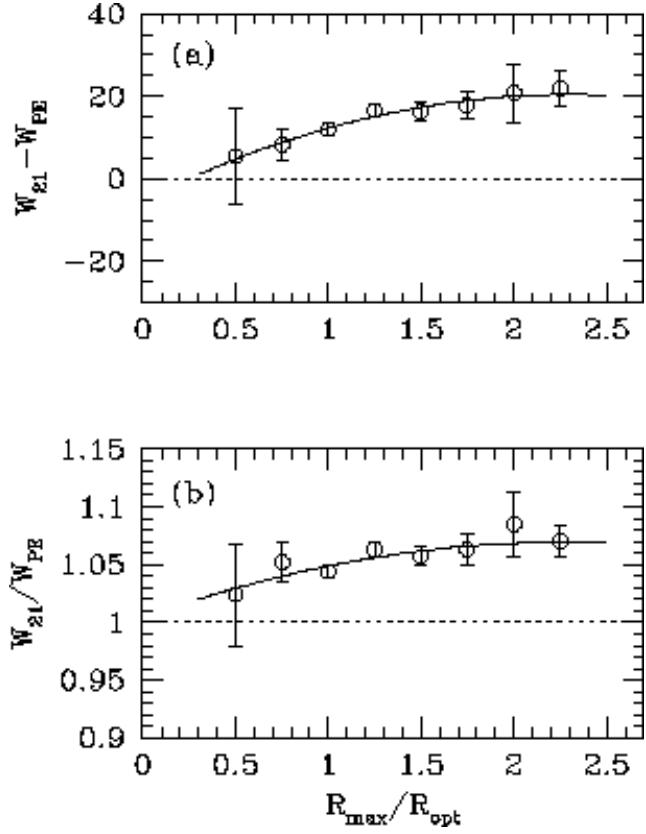}
\caption
{Comparison between \hi\ and Polyex widths as a function of RC
  extent. Bins and error bars are as in Figure \ref{opt_widths} (with
  a bin separation of 0.25 \ropt/\rd). The analytical expressions of the
  fits (solid lines) are listed in Table \ref{t_wconv}.
\label{radio_widths}}
\end{figure}

\begin{figure}
\epsscale{1.0}
\plotone{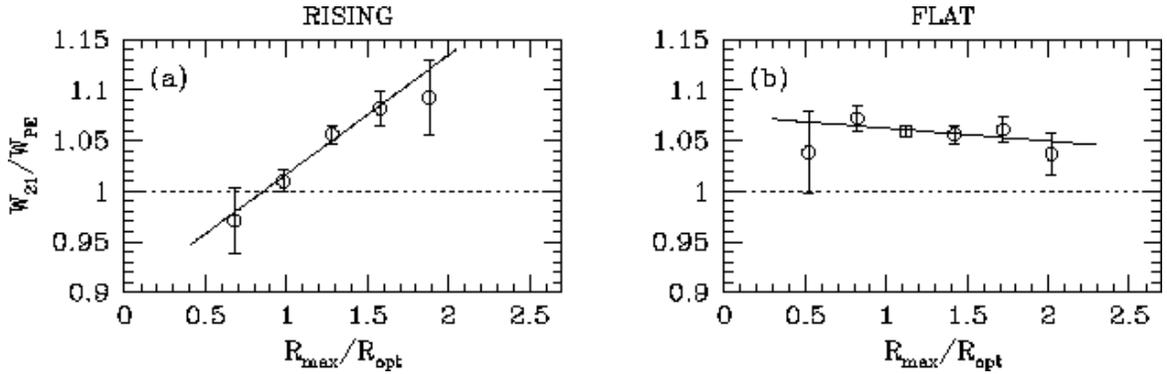}
\caption
{The ratio between \hi\ and Polyex widths is shown here
  separately for galaxies with rising (a) and flat (b) RCs. $\Theta$
  is the slope of the Polyex fit at the optical radius (whereas
  $\theta$ in Fig.~\ref{widths_sim} was defined as the slope of the
  fit between 0.5 and 1.0 \ropt); its units are \kms\ arcsec$^{-1}$. 
  Panels (a) and (b) show the results
  for 289 RCs with $\Theta >0.5$ and 566 RCs with $-0.5 < \Theta \leq 0.5$,
  respectively. The analytical expressions of the fits (solid lines)
  can be found in Table \ref{t_wconv}.
\label{awph_wph_slp}}
\end{figure}

\begin{figure}
\epsscale{0.55}
\plotone{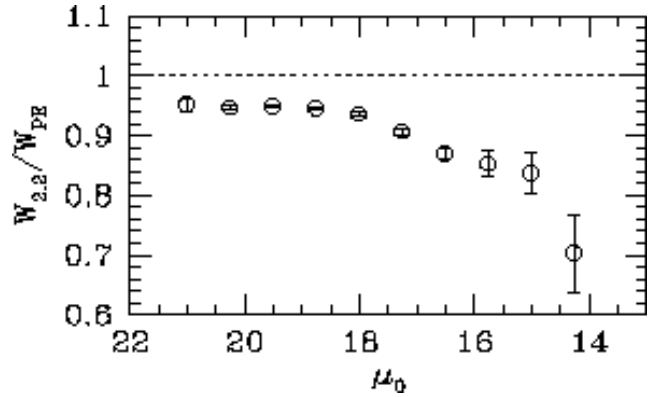}
\caption
{Dependence of the \wcou/\wpe\ RC width ratio on \iband\ disk central
  SB. Data points and error bars represent running averages and
  errors on the mean, respectively. 
\label{widths_sb}}
\end{figure}

\begin{figure}
\epsscale{0.55}
\plotone{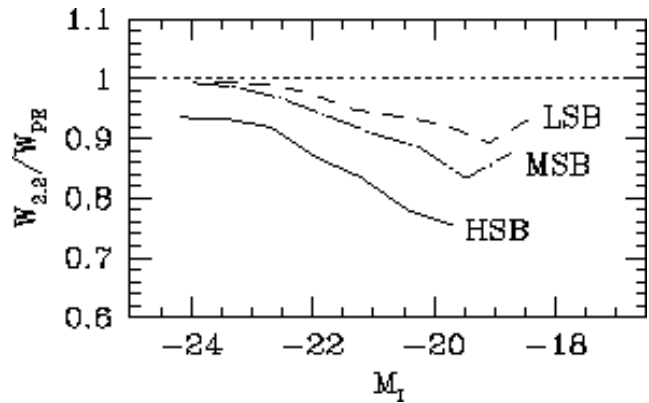}
\caption
{The \wcou/\wpe\ width ratio is plotted here as a function of the \iband\
  absolute magnitude. The lines indicate the values of running
  averages (symbols and error bars have been omitted for clarity) for
  galaxies with low (dashed), medium (dot-dashed) and high (solid) central
  surface brightness, as labeled. The division in SB classes is the
  same adopted in Fig.~\ref{opt_widths_sb}. 
\label{widths_msb}}
\end{figure}


\clearpage
\begin{deluxetable}{ll}
\tabletypesize{\scriptsize}
\tablecaption{Definitions of Radii and Velocity Widths Used in This Work \label{t_definitions}}
\tablewidth{0pt}
\tablehead{
\colhead{Symbol} & \colhead{Definition}
}
\startdata
  \rd      &  Scale length of the exponential disk from the \iband\ profile \\
  \rpe     &  Scale length of the exponential part of the Polyex fit to the RC \\
  \rmax    &  RC extent (i.e., maximum radius up to which the folded RC is sampled) \\
  \ropt    &  Radius encompassing 83\% of the total \iband\ light \\
  \rtw     &  Isophotal radius measured at an \iband\ surface magnitude of 23.5 mag arcsec$^{-2}$ \\
           &  \\
  \whis    &  Difference between the 90th and 10th percentile points of the RC velocity histogram \\
  \wpe     &  Twice the velocity at \ropt, measured from a Polyex fit to the RC \\
  \wcou    &  Twice the velocity at 2.15 \rd, measured from a Polyex fit to the RC \\
  \whi     &  Velocity width measured at the 50\% level of peak intensity on each side of the HI profile \\
\enddata 
\tablecomments{Throughout this work, radii are expressed in arcsec and
  velocity widths in \kms. The corrections applied to optical and
  radio widths are discussed in \S \ref{s_widths}.}
\end{deluxetable}

\begin{deluxetable}{rlc}
\tabletypesize{\scriptsize}
\tablecaption{Width Conversion Relations \label{t_wconv}}
\tablewidth{0pt}
\tablehead{
\colhead{$y$}&\colhead{$y = f$($x$=\rmax /\ropt)}&\colhead{RCs}
}
\startdata
\whi $-$\wpe    & = $-$5.21 \ +22.46 $x$ \ $-$4.88 $x^2$   & all \\
\whi /\wpe      & = \phd 1.004 \ +0.058 $x$ \ $-$0.012 $x^2$    & all \\
\whi /\wpe      & = \phd 0.899 \ +0.118 $x$                     & rising \\
\whi /\wpe      & = \phd 1.075 \ $-$0.013 $x$                   & flat \\
\enddata 
\end{deluxetable}

\end{document}